\documentclass{optica-article}

\journal{optcon}
\articletype{Research Article}
\begin{document}

\title{Large Quality Factor Enhancement Based on Cascaded Uniform Lithium Niobate Bichromatic Photonic Crystal Cavities}

\author{Rui Ge,\authormark{1} Xiongshuo Yan,\authormark{1} Zhaokang Liang, \authormark{1} Hao Li,\authormark{1} Jiangwei Wu, \authormark{1} Xiangmin Liu,\authormark{1} Yuping Chen,\authormark{1,*} and Xianfeng Chen, \authormark{1,2,3} }

\address{\authormark{1}State Key Laboratory of Advanced Optical Communication Systems and Networks, School of Physics and Astronomy, Shanghai Jiao Tong University, 800 Dongchuan Road, Shanghai 200240, China\\
\authormark{2}Shanghai Research Center for Quantum Sciences, Shanghai 201315, China\\
\authormark{3}Collaborative Innovation Center of Light Manipulations and Applications, Shandong Normal University}

\email{\authormark{*}Corresponding author: ypchen@sjtu.edu.cn} %% email address is required; see note below about the corresponding author designation



\begin{abstract}
 In this paper, by cascading several bichromatic photonic crystals we demonstrate that  the quality factor can be much larger compared with that in an isolated cavity without increasing the total size of the device. We take lithium niobate photonic crystal as an example to illustrate that the simulated quality factor of the cascaded cavity can attain 10$^{5}$ with a 70° slant angle, which is an order of magnitude larger than that in isolated cavity. The device can be fabricated easily by current etching technique for lithium niobate. We have fabricated the proposed device experimentally including holes with ~70° slant angle. This work is expected to provide guidance to the design of photonic crystal cavity with high-quality factor.
\end{abstract}

%%%%%%%%%%%%%%%%%%%%%%%%%%  body  %%%%%%%%%%%%%%%%%%%%%%%%%%

Photonic crystal cavity can trap photons in sub-wavelength scale and becomes a hot spot recently. High-quality factor photonic crystal cavity is beneficial for strong nonlinear effect \cite{li2019high}, ultra-sensitive sensor \cite{wang2019single}, and high-transmission efficiency notch filter \cite{brunetti2019ultra}. Previous works have proposed many approaches to achieve the high-quality factor photonic crystal cavity. The end-hole shifted cavity is proved to show a high quality factor \cite{akahane2003high}. Being derived from such mechanism, mode-gap photonic crystal cavity is then proposed \cite{zhang2015review,tanaka2008design,zhou2022photonic}. The lattice constant or radius of holes of the photonic crystal can be gradually adjusted along the line defect and the cavity is arranged in a heterostructure type \cite{zhang2015review}. Researchers found that the light in a mode-gap photonic crystal cavity can be confined with Gaussian shape with the ignorable energy within the light cone in the Fourier transformation spectrum, yielding an ultra-high quality factor \cite{tanaka2008design}. Except for these methods, inverse design \cite{asano2019iterative} and visualization of leaky components \cite{nakamura2016improvement} are also effective. \par
Bichromatic photonic crystal cavity with Aubry-André-Harper (AAH) quasi-periodic potential only require two incommensurable lattices without sophisticated design which is an excellent platform to achieve high-quality factor photonic crystal cavity and can achieve the largest theoretical quality factor among various kinds of designs \cite{alpeggiani2015effective}. Another advantage of bichromatic cavity lies in that it supports the topological edge state in the synthetic dimension \cite{alpeggiani2019topological}. Bichromatic photonic crystal cavity was already used in third harmonic generation \cite{simbula2017realization}, photon comb generation \cite{combrie2017comb,clementi2020selective}, opto-mechanical coupling \cite{ghorbel2019optomechanical}, quantum dot coupling \cite{rickert2020mode}, and optical parametric oscillator \cite{marty2021photonic}. The relative experiment proved that the quality factor obtained in experiment has an upper boundary because as the intrinsic quality factor of the device becomes large the coulping quality factor becomes dominant \cite{simbula2017realization}. Consequently, one vital goal of designing a cavity lies in that obtaining a larger quality factor but with the minimum number of holes. Previous works have verified that to further improve the quality factor of bichromatic photonic crystal, the ratio of two lattice constants can be adjusted and gradually approaches 1 (but not 1) \cite{simbula2017realization}, which will inevitably enlarge the size of the whole photonic crystal. The radius of lattice in the line defect can also becomes smaller to enhance the quality factor, which is limited by the current fabrication process. Other methods include optimizing positions of some holes which require lots of extra simulations \cite{minkov2017photonic} or pushing defect holes outer for a small distance \cite{kuruma2020strong}. Methods to further improve the quality factor are required. \par
In this paper, we show that the cascaded bichromatic photonic crystal cavity can exhibit a higher quality factor compared with the one in isolated bichromatic photonic crystal cavity. We explain this phenomenon by analyzing the spatial Fourier transformation spectrum for a cascaded and an isolated birchromatic cavity. The cascaded bichromatic photonic crystal cavity which is constructed by arraying bichromatic lattice in vertical direction can effectively utilize the freedom of $y$ dimension. In previous works our group has already theoretically or experimentally demonstrated the beam splitter \cite{duan2016broadband}, logic gate \cite{lu2019all}, observable second harmonic generation \cite{jiang2018nonlinear}, opto-mechanical coulping \cite{jiang2020high} and valley waveguide \cite{ge2021broadband} with lithium niobate (LN) photonic crystal. We then fabricated the LN photonic crystal to demonstrate that our simulation model is compatible with current etching process for LN.

\begin{figure}[ht]
\centering
\includegraphics[width=\linewidth]{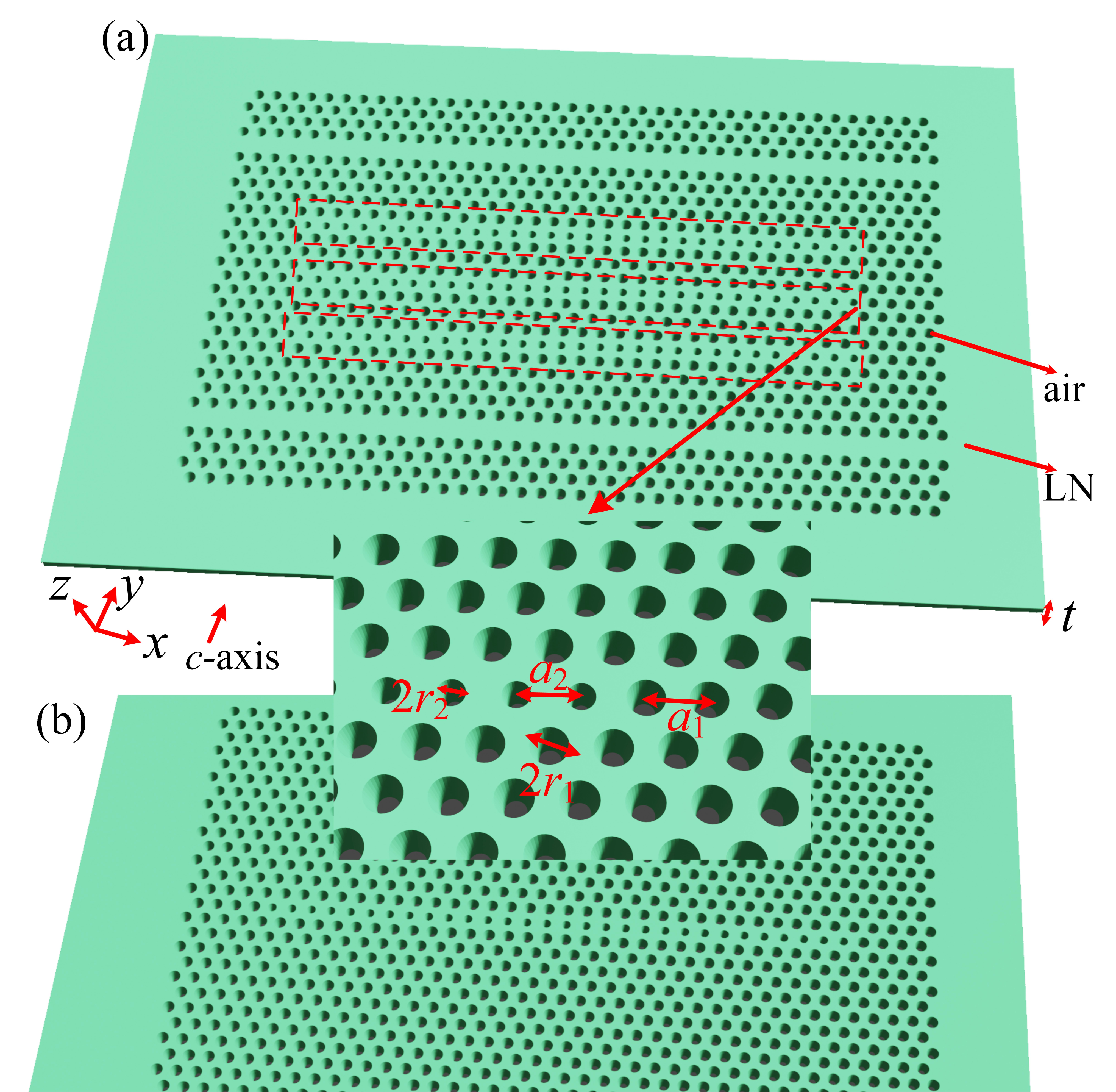}
\caption{(a) Model of cascaded bichromatic photonic crystal cavities with inset: enlarged view of the right part of the device. (b) Model of isolated bichromatic photonic crystal cavity.}
\label{fig:1}
\end{figure}

Firstly we consider the standard bichromatic photonic crystal cavity which is constructed by tiling two kinds of holes with different lattice constants and radii. Previous work has already demonstrated that the design shows Gaussian shape energy distribution which leads to a high-quality factor \cite{alpeggiani2015effective}. Here we array the standard bichromatic photonic crystal in $y$ direction as shown in three-dimensions (3D) model in Fig. \ref{fig:1}(a) and the isolated cavity is also shown in Fig.\ref{fig:1}(b). The dashed red boxes indicate the line defects. The radius of bulk holes is set to $r_{1}$=260 nm, and the radius of small holes in the line defect region is set to $r_{2}$=110 nm. The thickness of the suspended slab is 300 nm. The material is chosen as LN with the optical axis pointing to $y$ direction. Consequently, the refractive index of the slab is $n_{x}$=2.21, $n_{y}$=2.13 and $n_{z}$=2.21 \cite{li2015design}.
When etching the holes in slab, there will always be a 70° slant angle for each air hole \cite{qi2020integrated} and we also taken that into consideration, i.e., the shape of a hole is a truncated cone with a 70° slant angle instead of a cylinder. The lattice constant of bulk region is $a_{1}$ =650 nm, and in the line defect region the lattice constant is $a_{2}$=650 nm×29/30=628.3 nm. The detailed design principle can be referred to Ref. \cite{alpeggiani2015effective} . Then we array these bichromatic photonic crystal cavities and introduce two extra line defects. The whole device is like an add-drop filter. The isolated bichromatic photonic crystal cavity has no add or drop channels. In Fig.\ref{fig:1} (a) the source is placed at the upper-left port and the monitor is placed at the right-lower port. In Fig.\ref{fig:1} (b) the source is placed on the left side and the monitor is placed on the right side. \par

\begin{figure}[ht]
\centering
\includegraphics[width=\linewidth]{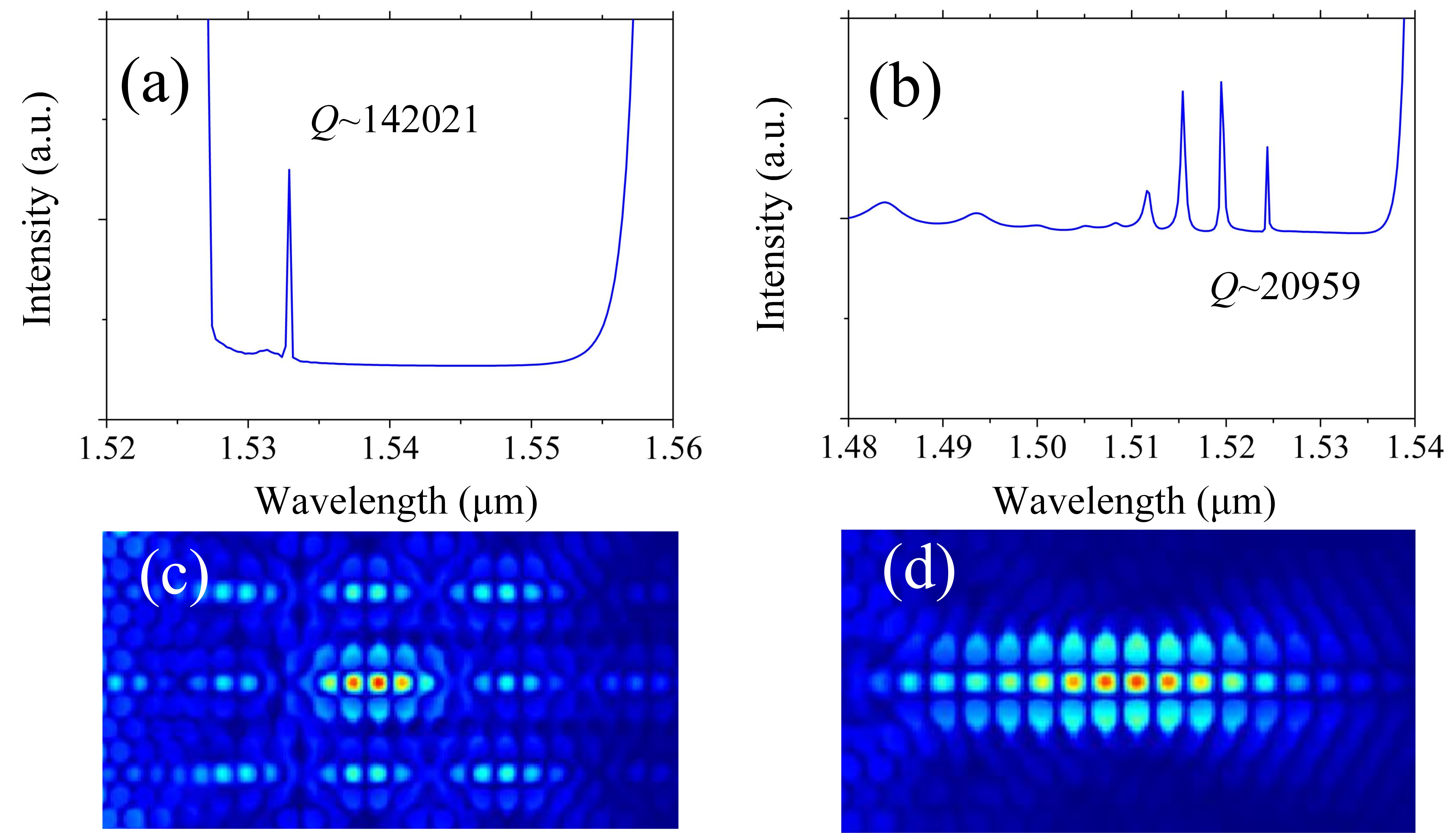}
\caption{Transmission spectra of the (a) cascaded and (b) isolated bichromatic photonic crystal cavities. Mode profiles of (c) cascaded and (d) isolated bichromatic photonic crystal cavities. }
\label{fig:2}
\end{figure}

We simulated the transmission spectrum of the designed structure as shown in Fig.\ref{fig:2} (a). The device support transverse electric (TE) mode band gap. For the cascaded bichromatic photonic crystal we find the transmission spectrum shows one resonant mode with a quality factor of ~1.4×10$^{5}$, while for the isolated bichromatic photonic crystal in Fig.\ref{fig:2} (b) there are several resonant modes which form a “comb” \cite{combrie2017comb} . The right-most resonant mode shows a quality factor of 2×10$^{4}$, and other resonant modes are found to have the quality factors that are lower than 10$^{4}$. Consequently, we mainly concentrate on the mode with the largest quality factor in the below discussion. The mode profiles for the cascaded and isolated bichromatic photonic crystal cavities are also plotted in Fig.\ref{fig:2}(c)-\ref{fig:2}(d). The results indicate that the energies are tightly confined in the central region. 

\begin{figure}[ht]
\centering
\includegraphics[width=\linewidth]{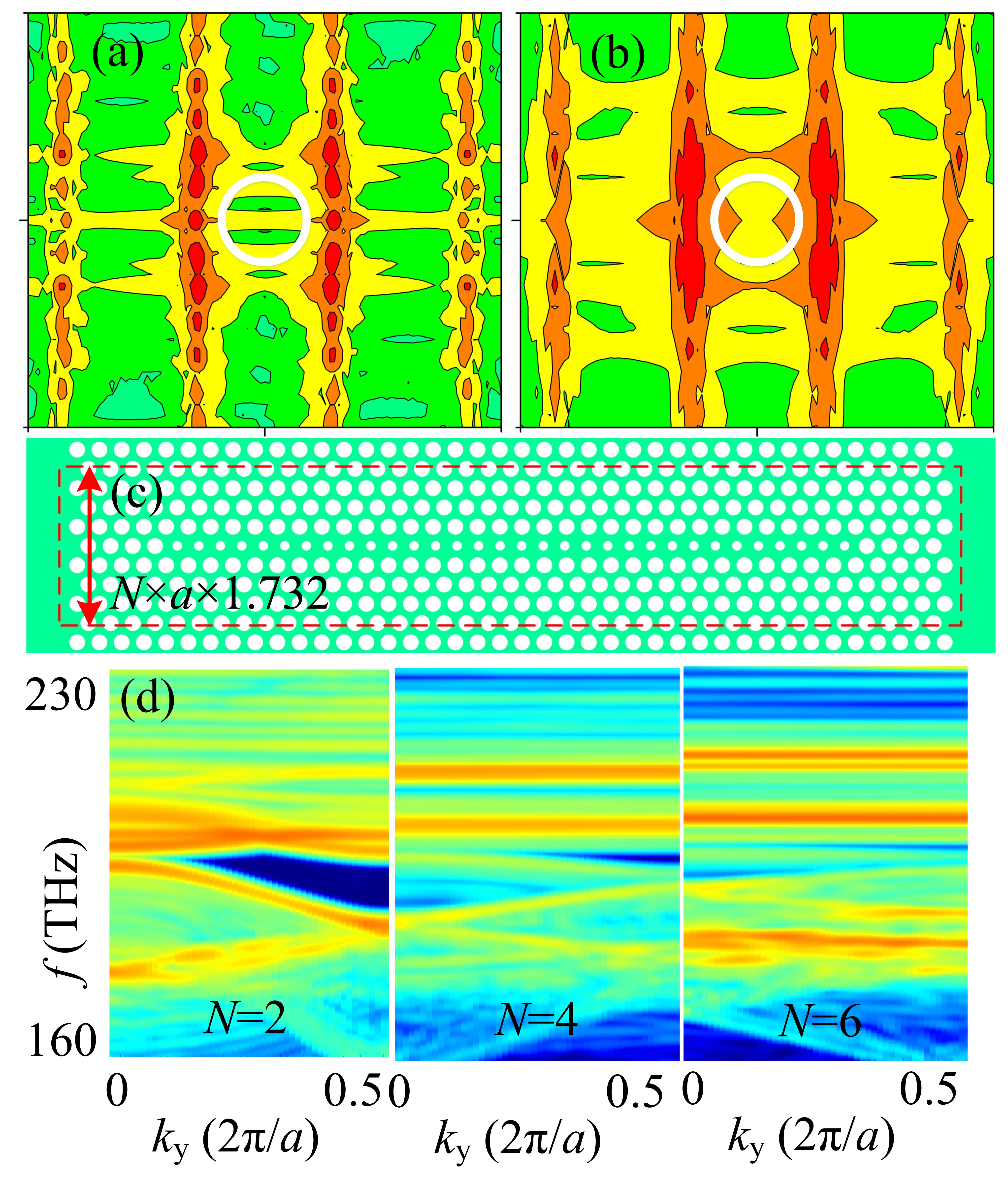}
\caption{Spatial Fourier transform spectra for the (a) cascaded and (b) isolated bichromatic photonic crystal. (c) Model of isolated bichromatic photonic crystal with Bloch boundary conditions. (d) Simulated projected band diagrams.}
\label{fig:3}
\end{figure}

The simulation results indicate that the quality factor of cascaded bichromatic photonic crystal is almost an order of magnitude larger than that of isolated one and the cascaded bichromatic photonic crystal shows less resonant modes.\par
Aiming at the explanation for why the quality factor of cascaded photonic crystal cavities is larger, we plot the spatial Fourier transform spectra of the cascaded and isolated bichromatic photonic crystal cavities based on the $|H_{z}|$ field at $z$=0 plane as shown in Fig. \ref{fig:3}(a)-\ref{fig:3}(b) and different colors represent different logarithm amplitudes of the Fourier transform results. The light line is labeled by the white solid line and can be given by the formula $\omega$=$ck$ \cite{alpeggiani2015effective} , where $\omega$ is the frequency, $c$ is the light speed and $k$ is the wave vector. The energy within the light line of the cascaded bichromatic photonic crystal is far lower than that of the isolated one from these figures. The cascaded photonic crystal cavities in our work construct a coupled resonant optical waveguide (CROW). Consequently, we utilize the CROW tight binding approximation (TBA) for analyzing the quality factor of our device \cite{jagerska2009radiation,ma2013tight} . 
According to TBA, the quality factor of a cavity is determined by the relative Fourier transform value within the light line region, and the lower energy within the light line in Fig. \ref{fig:3}(a) shows that the quality factor of cascaded bichromatic cavity is larger. The Fourier transform CROW field can be expressed by terms of isolated cavity fields $E_\Omega$ and $H_\Omega$ \cite{ma2013tight} :

\begin{equation}
\begin{split}   
|\Tilde{E}_{k,i}(q_x,q_y)|^2=  |E_0|^2 (\frac{2\pi}{D})\sum_{m=-\infty}^{\infty} \sum_{n=-\infty}^{\infty} \\ \delta(q_x-k-\frac{2\pi n}{D})\times
|\Tilde{E}_{\Omega,i}(q_x,q_y)|^2
\end{split}
\label{eq:4}
\end{equation}

\begin{equation}
\begin{split}   
|\Tilde{H}_{k,i}(q_x,q_y)|^2=  |H_0|^2 (\frac{2\pi}{D})\sum_{m=-\infty}^{\infty} \sum_{n=-\infty}^{\infty} \\ \delta(q_x-k-\frac{2\pi n}{D})\times |\Tilde{H}_{\Omega,i}(q_x,q_y)|^2
\end{split}
\label{eq:5}
\end{equation}

Where $D$ is the center-to-center distance of adjacent line defects and the sum over $m$ is a sum over unit cells. $\Tilde{E}_{k,i} (q_x, q_y)$ and $\Tilde{H}_{k,i} (q_x, q_y)$ are Fourier transforms of the field. $E_{0}$ and $H_{0}$ are normalization constants. $q_{x}$ and $q_{y}$ are coordinates in Fourier space.

The above discussions illustrate that the Fourier transform field of CROW can be deduced from that of the isolated cavity. For a CROW with infinite length, the Fourier transform field is the product of Fourier transform field of the isolated cavity and some delta functions. For a CROW with finite length, the Fourier transform field is the product of Fourier transform field of the isolated cavity and some Sinc functions \cite{jagerska2009radiation,ma2013tight}. The principle can be clearly seen in Fig. \ref{fig:3}(a)-\ref{fig:3}(b): if we periodically sample the field with a period of $2\pi/D$ in Fig. \ref{fig:3}(b) we can obtain Fig. \ref{fig:3}(a). After sampling, the energy in some region are filtered out periodically. Obviously most energy within the light line in Fig. \ref{fig:3}(b) mainly concentrates on the two fan-shaped lobes. However, the energy out of the light line disperses uniformly in two ribbons. After sampling, most energy in the lobes vanished but there is still much energy in the ribbon. Consequently, the relative energy within the light line in Fig. \ref{fig:3}(a) is lower mainly owing to that the implementation of sampling filtered out most energy within the light line but it has limited influence on the energy out of the light line. 

\begin{figure}[ht]
\centering
\includegraphics[width=\linewidth]{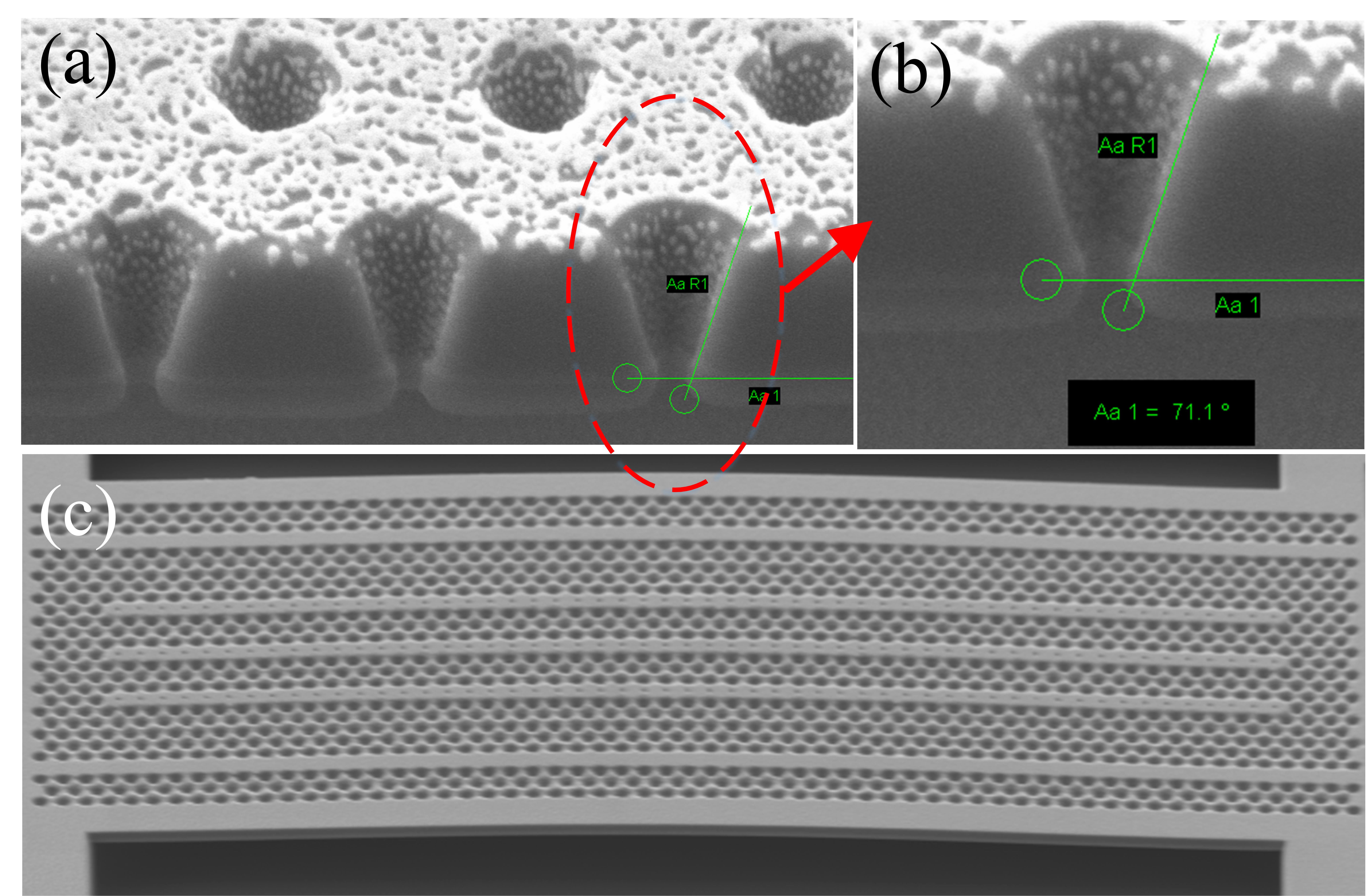}
\caption{(a) Cross view of the dry-etched holes with slant angle of 71.1°. (b) Enlarged view of (a). (c) SEM graph of fabricated bichromatic cavity.}
\label{fig:4}
\end{figure}

\begin{figure}[ht]
\centering
\includegraphics[width=\linewidth]{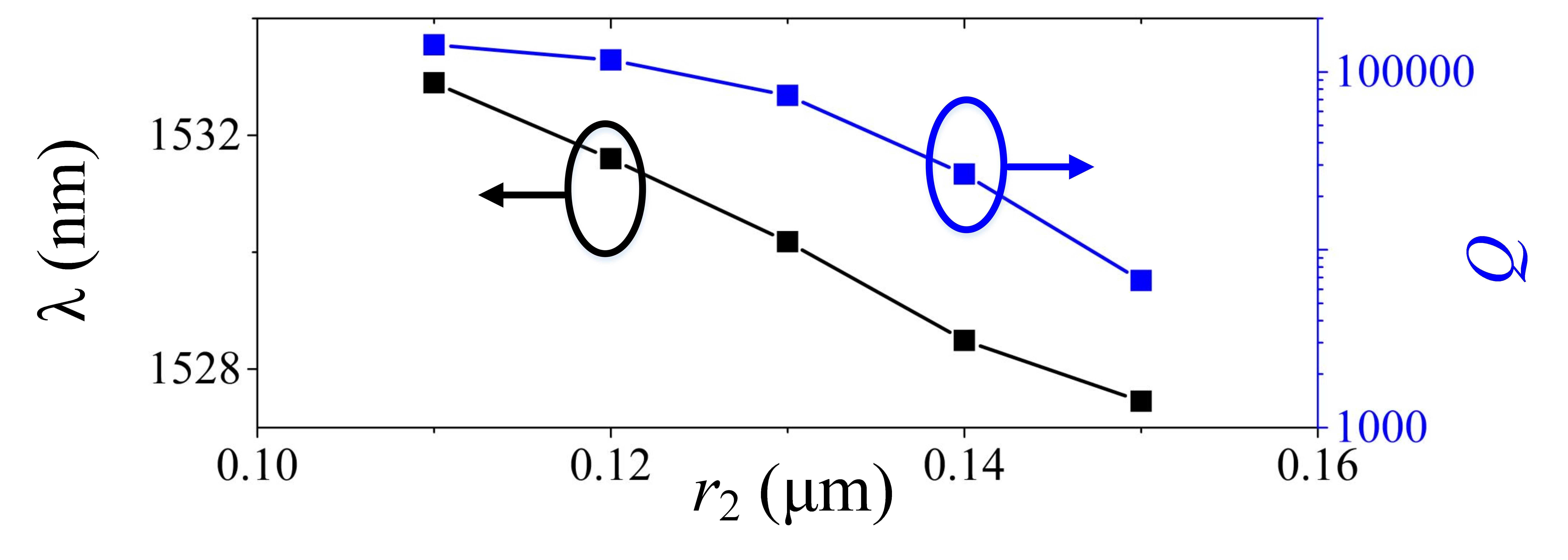}
\caption{Dependence of $r_{2}$ on resonant wavelength and quality factor of the device.}
\label{fig:5}
\end{figure}

Then we attempt to figure out why the number of resonant modes in the cascaded cavities is less than that in the isolated cavity.  From the view of the dispersion relationship, the formation of comb results from the existence of ultra-flat bands. Once the Bloch boundary conditions are applied to the bichromatic cavity to obtain the dispersion relationship, the resonance modes in Fig. \ref{fig:2}(b) are actually the ultra-flat bands in the projected band diagram \cite{alpeggiani2019topological}. When the $\beta$ is small, there are several bands in the band gap region and only one or two of them are flat. As $\beta$ becomes larger (approaching 1 but not 1), the bands in higher frequencies become flat and they will form localized resonant modes \cite{alpeggiani2019topological}. These flat bands are supported by the AAH potential and once $\beta$ is small, the bichromatic cavity shows much difference with standard AAH lattice. Here we simulated the projected band diagram for our structure in Fig. \ref{fig:3}(c). The simulated model is the isolated bichromatic photonic crystal cavity, and we set the upper and lower sides to Bloch boundary conditions accompanied by perfect matched layer boundary conditions on the left and right sides. It should be noted that the direction of Bloch boundary condition is different from the one in Ref. \cite{alpeggiani2019topological}. The height of the unit cell is $1.732Na$ and the width of the unit cell is large enough. To rule out the influence of transverse magnetic mode we set the radius of bulk holes to 220 nm, the slant angle of all holes to 90°, and the radius of small holes to 80 nm. Other geometry parameters are the same as those in Fig. \ref{fig:1}(a). The projected band diagrams of TE modes with $N$=2, $N$=4 and $N$=6 are plotted in Fig. \ref{fig:3}(d), respectively. The projected band diagram of $N$=6 shows 3 flat bands while it decreases to 2 when $N$=4. As for the case of $N$=2, only one flat band can be found. In general, the result of N=6 is more like the case of the isolated cavity and the result of $N$=2 is more like the case of the cascaded cavities. The above results indicate that as $N$ gradually decreases from a large number, the flat bands gradually disappeared. It can be found that the band in the figure of $N$=2 shows some similarities with the that of small $\beta$ in Ref. \cite{alpeggiani2019topological} where only one band is localized while other bands are not flat. Consequently, the disappearance of resonant modes results from the weakening of AAH potential. 
We must underline that cascading bichromatic cavities cannot necessarily enhance the quality factor compared with that in the isolated one. Owing to that we arrange these cavities in the form of add-drop filter, only the light supported by the waveguide band can successfully enter the device and there will be mode mismatch between waveguide modes and CROW modes. As for the isolated cavity, the light with frequency within the band gap region can easily trigger the resonance. This phenomenon is verified in Fig. \ref{fig:2}(a)-\ref{fig:2}(b) where it can be found that the region with low transmission efficiency in Fig. \ref{fig:2}(b) is far larger than that in Fig. \ref{fig:2}(a). What’s worse, being set to unsuitable parameters the resonant mode will submerge in the bulk transmission region and cannot be found. In addition, once $D$ or the number of arrayed units is not chosen correctly, sampling cannot filter out the energy within the light line effectively, which will make the quality factor remain unchanged. 

Finally we fabricated the proposed device to verify that our simulated model is compatible with the current etching technique. Our structure is fabricated based on the $x$-cut lithium niobate on insulator (LNOI). The 410 nm thick amorphous silicon is deposited as the mask, and then the device is patterned with ZEP520 positive resist via electron beam lithography technique. The pattern is transferred to the mask layer by reactive ion etching and then transferred to LN layer with Ar+ plasma etching \cite{li2019high}. After removing the resist and mask the silica substrate is removed by hydrofluoric acid. We etched some holes and cut them by focus ion beam along the center line of the hole, the slant angle of 71.1° can be observed as shown in scanning electron microscope (SEM) graph in Fig. \ref{fig:4}(a)-\ref{fig:4}(b). The bichromatic photonic crystal is then fabricated as shown in Fig. \ref{fig:4}(c). Previous works have already proved that as the radius of defect holes increases the quality factor will decrease for the isolated cavity \cite{simbula2017realization}. Here we simulate the quality factor and resonant wavelength for the cascaded bichromatic cavity and a similar trend can be obtained as shown in Fig. \ref{fig:5}. The quality factor and resonant wavelength both decrease as the $r_{2}$ increases.

In conclusion, we proposed a high-quality factor cavity by cascading bichromatic photonic crystals. Our simulation model takes a 70° slant angle into consideration which is compatible with current etching process of LN. The design do not require numerous iterative simulations and is generally simple. The physical mechanism of the enhanced quality factor is also illustrated. This work is expected to provide guidance to the design of high-quality factor cavity with other materials and the construction recipe of the cavity is not limited by the size, shape, and thickness of the holes.

\begin{backmatter}
\bmsection{Funding} This work was supported by the National Natural Science Foundation of China (Grant Nos. 91950107, and 12134009), the National Key R\&D Program of China (Grant Nos. 2019YFB2203501), Shanghai Municipal Science and Technology Major Project (2019SHZDZX01-ZX06), and SJTU No. 21X010200828.

\bmsection{Acknowledgments} We thank the Center for Advanced
Electronic Materials and Devices (AEMD) of Shanghai Jiao
Tong University for the assistance of device fabrication. We thank Wang Ying at AEMD for the guidance of fabrication.
\bmsection{Disclosures} The authors declare no conflicts of interest.

\bmsection{Data Availability Statement} Data underlying the results presented in this paper are not publicly available at this time but may be obtained from the authors upon reasonable request.

\end{backmatter}

%%%%%%%%%%%%%%%%%%%%%%% References %%%%%%%%%%%%%%%%%%%%%%%%%

%%%%%%%%%% If using BibTeX:
\bibliography{sample}

%%%%%%%%%% If preparing manually:
% \newcommand{\enquote}[1]{``#1''}

% \bibitem{Zhang:14}
% Y.~Zhang, S.~Qiao, L.~Sun, Q.~W. Shi, W.~Huang, L.~Li, and Z.~Yang,
%   \enquote{Photoinduced active terahertz metamaterials with nanostructured
%   vanadium dioxide film deposited by sol-gel method,}
%   {\protect\JournalTitle{Optics Express}} \textbf{22}, 11070--11078 (2014).

% \bibitem{Optica}
% {Optica}, \enquote{{Optica Publishing Group},}
%   \url{http://www.opg.optica.org}.

% \bibitem{FORSTER2007}
% P.~Forster, V.~Ramaswamy, P.~Artaxo, T.~Bernsten, R.~Betts, D.~Fahey,
%   J.~Haywood, J.~Lean, D.~Lowe, G.~Myhre, J.~Nganga, R.~Prinn, G.~Raga,
%   M.~Schulz, and R.~V. Dorland, \enquote{Changes in atmospheric consituents and
%   in radiative forcing,} in \enquote{Climate Change 2007: The Physical Science
%   Basis. Contribution of Working Group 1 to the Fourth assesment report of
%   Intergovernmental Panel on Climate Change,}  S.~Solomon, D.~Qin, M.~Manning,
%   Z.~Chen, M.~Marquis, K.~B. Averyt, M.~Tignor, and H.~L. Miler, eds.
%   (Cambridge University Press, 2007).

% \end{thebibliography}

\end{document}